\documentclass[manuscript]{acmart}
\AtBeginDocument{%
  }

\renewcommand\footnotetextcopyrightpermission[1]{} 

\setcopyright{acmlicensed}
\copyrightyear{2026}
\acmYear{2026}
\acmDOI{}
\acmConference[CHI '26]{Data Literacy Workshop}{April 13-17, 2026}{Barcelona, Spain}

\begin{document}

\title{An Experiential Approach to AI Literacy}

\author{Aakanksha Khandwaha}
\affiliation{\institution{University of Waterloo}
\city{Waterloo}
\country{Canada}}
\email{aakanksha.khandwaha@uwaterloo.ca}

\author{Edith Law}
\affiliation{\institution{University of Waterloo}
\city{Waterloo}
\country{Canada}}
\email{edith.law@uwaterloo.ca}


\begin{abstract}
Despite AI tools becoming more prevalent and applicable to a variety of workplaces, workers consistently report uncertainty about where AI applies, what problems it can help solve, and how it fits into real workflows. In other words, there is a gap between `knowing' and `doing' when it comes to AI literacy. We propose an experiential form of AI literacy which integrates participant's daily experiences into the learning experience by brainstorming grounded AI use cases through storytelling. We introduce a novel pedagogical approach that helps individuals move away from abstract notions of AI towards practical knowledge of how AI would (or would not) work in different workflows, contexts, and situations. Through this approach, we anticipate two major outcomes: (1) enhanced AI literacy for stakeholders within a variety of work sectors and (2) concrete AI use cases developed through participatory design that are grounded in AI literacy and participant's expertise.
\end{abstract}

\keywords{AI Literacy, Experiential Learning, Participatory AI, Co-design}

\maketitle

\section{Introduction \& Background}

Despite AI-enabled tools becoming prevalent, implementation of AI solutions in workplaces such as healthcare, education, and government remains challenging. Workers often express uncertainty about where AI applies, what problems it can help solve, and how it fits into their pre-existing workflows  \cite{babashahi_ai_2024}. Many have little exposure to responsible-use examples, making it difficult to identify opportunities or articulate needs during co-development.

Underpinning these challenges is a broader gap in public perception and understanding of AI: many teams are unsure how AI systems work, what data they require, how to judge model performance, or what human oversight looks like in practice \cite{babashahi_ai_2024}. Without this foundation, staff may either overestimate AI’s capabilities or distrust it entirely. Compounding this is a noticeable apprehension that AI may replace workers.
Misconceptions about automation, job displacement, and “black box” decision-making create hesitation and resistance, even when tools are designed to relieve burden and strengthen decision-making. Decision-makers also require deeper capability in use-case evaluation—understanding how to determine whether AI is appropriate for a given challenge, what workflow changes are necessary, and how to ensure alignment with safety, privacy and organizational priorities \cite{pereira_systematic_2023}.

AI literacy has been a rising topic of interest aimed at addressing this gap. Scholars have varying definitions, but the most well-established one outlines it as “competencies that enables individuals to critically evaluate AI technologies, communicate and collaborate effectively with AI, and use AI as a tool online, at home and in the workplace” \cite{long_co-designing_2021}. It intersects with digital literacy, data literacy and competency-based learning models \cite{collyer-hoar_experts_2025, long_co-designing_2021,mannila_co-designing_2024}. 
Recently, there has been an emphasis on designing AI literacy with the end user’s day-to-day life and knowledge in mind, referred to as the “stakeholder-first” approach \cite{dominguez_figaredo_responsible_2023}. Many AI literacy frameworks and design considerations also point to the importance of prioritizing the population that the AI literacy tool(s) are being developed for \cite{long_what_2020, ng_conceptualizing_2021,xie_exploring_2025}. To this end, researchers have co-designed AI literacy materials and tools with a variety of stakeholders, such as teachers \cite{dominguez_figaredo_responsible_2023, laupichler_artificial_2022,long_what_2020}, child-computer interaction experts \cite{baguley_more_2022} and museum workers \cite{long_co-designing_2021}, or developed design toolkits to help them envision how AI can be used \cite{sadeghian_workai_2025, yildirim_creating_2023,smith_codesigning_2025, bhat_designing_2024}.

Despite these advances, there is still a gap between `knowing' and `doing' when it comes to AI literacy, i.e., AI education often does not translate to understanding how to make decisions about AI adoption in particular work contexts, especially for non-technical audiences. Experiential learning aims to foster a deeper understanding of the subject at hand through reflection and personal experience \cite{kolb_experiential_2014}. Within AI education, experiential learning has been shown to enhance AI literacy among K-12 and post-secondary students \cite{gnoth_supporting_2025, hsu_is_2021, forster_building_2024}. However, this work focuses on making the workshops themselves be more experiential through interactive activities, rather than incorporating AI literacy within participants' daily lives. A similar approach has been applied to other types of data literacies, such as visualization literacy, where participants learn to visualize personal data such as daily activities or personal interests, for a more integrated and reflective educational experience \cite{dignazio_creative_2018}.

In this paper, we propose a unique experiential approach to AI literacy, where stakeholders from diverse workplaces learn and understand AI capabilities via brainstorming use cases grounded in their personal experiences.  Through storytelling, we aim to shift understanding from abstract concepts toward practical knowledge of how AI may or may not work within specific workflows, contexts, and situations. We anticipate that this will lead to enhanced AI literacy among people across diverse work sectors.

This experiential method of learning will also lead to better participatory AI practices, where stakeholders are involved throughout the design process \cite{birhane_power_2022}. Specifically, we focus on co-designing AI, which focuses on knowledge sharing to foster a collective understanding between stakeholders \cite{smith_codesigning_2025}. Involving stakeholders who are AI literate is paramount for co-designing AI, which our approach facilitates. However, more importantly, our method encourages them to consider AI within the context of their own lives, leading to more insightful contributions towards its design and development.
\section{Our Approach}

\begin{figure}[!h!t]
    \centering
    \includegraphics[width=1\linewidth]{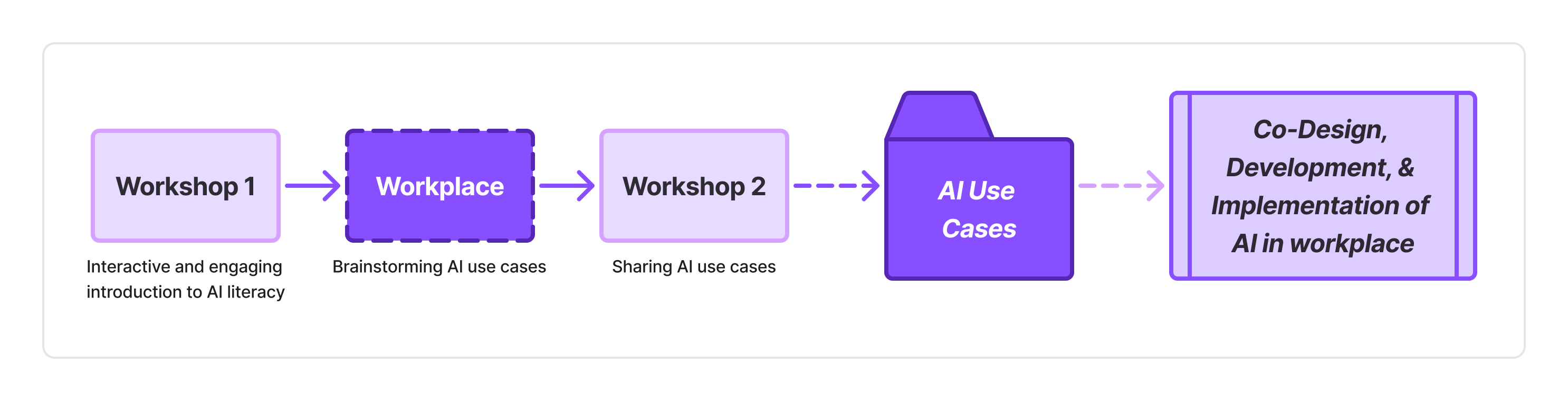}
    \caption{Our approach to AI literacy, which grounds participant's understanding of AI in their work context through the development of AI use cases.}
    \Description{A flowchart depicting our approach to AI literacy, starting with Workshop 1, leading to participant's workplace where they brainstorm AI use cases, leading to Workshop 2 where they share use cases, resulting in AI use cases which can be developed and implemented in workplaces.}
    \label{fig:ai-literacy}
\end{figure}

In this position paper, we propose an \textbf{experiential approach to AI literacy} which grounds participants' understanding of AI within their work contexts to better facilitate deeper conversations around AI capabilities and limitations. This involves \textbf{three phases} (also outlined in Figure \ref{fig:ai-literacy}): (1) \textit{an initial workshop}, where participants are given an interactive introduction to AI, (2) \textit{an experiential component}, where participants reflect on and brainstorm AI use cases within their workplace over several weeks, (3) \textit{a sharing workshop}, where participants share the AI use cases that they discovered, including opportunities, limitations, and ethical considerations with their particular scenario. This will result in a set of AI use cases that are practical, relevant, and feasible within participant's work sectors. Each workshop should include participants within similar work contexts (e.g., nurses in emergency departments, elementary school teachers, municipal government workers) which will allow us to go in depth about the topics, use cases, and pitfalls of AI. 

The first workshop should be an engaging, interactive session which forms the participant's understanding of AI. This should draw from the extensive literature in AI literacy that helps non-technical audiences gain an understanding of how AI works, how AI is used, how to evaluate AI, and its limitations \cite{ng_conceptualizing_2021, long_what_2020}. We believe it is important that this workshop stays relevant and reliable over a long period of time, while allowing for concrete additions that help contextualize current AI use, given that public perception of AI constantly changes. Thus, we want to focus on high-level, abstract understandings of AI, (e.g., AI helps discover, predict, identify, or generate \cite{yildirim_creating_2023}) rather than diving into the latest AI tools available. From this, we can provide examples of specific AI that participants may be familiar with (e.g., chatbots, social media recommendation algorithms, grammar and spellcheck) to help situate them within their own experiences. Ultimately, the goal of this workshop is to provide them with a broad but personalized understanding of AI, so that they can better interpret past experiences with AI and brainstorm potential future experiences.

After the first workshop, participants should have a baseline understanding of AI which would allow them to describe current and/or imagined AI capabilities in their own work contexts. They will be provided with some design tools and activities (similar to WorkAI \cite{sadeghian_workai_2025}, Smith et al.'s card-based co-design toolkit \cite{smith_codesigning_2025}, or the AI brainstorming kit \cite{yildirim_creating_2023}) to help facilitate this process. This part of the learning process would approximately take 2-4 weeks to provide them with the time to reflect and deliberate on their daily interactions with technology and challenges they experience with work, and whether/how AI could help alleviate them. When participants have time, either during or after their workday, they can complete brainstorming activities that would help them flesh out their challenges, experiences, and how they imagine AI integrating into their workflow. It is important to note that participants' ability to brainstorm may be limited by the demands of their work or home life, so the activities should be short, easy, and enjoyable. The focus is on integrating reflective practices into participant's daily routines, allowing them to identify challenges and envision AI use cases aligned with their needs.

Participants will join a second workshop after the brainstorming phase, where they will share their use cases through detailed stories as rationale, based on challenge(s) they observed or experienced in their workplace. This stage emphasizes AI ethics, since participants can concretely understand the limitations of AI through critically evaluating a given use case for fairness, privacy, accountability, transparency, etc. This will lead to deeper discussions about AI, resulting in a practical understanding of AI based on the participant's shared knowledge and experiences. This will also lead to an iterative refinement process of participants' use cases, which will include details on workflow integration, required resources, and potential challenges.

As mentioned previously, this approach also supports the participatory design of AI through the development of AI use cases. At the end, these workshops and brainstorming activities should generate a set of AI use cases that is grounded in a nuanced understanding of AI capabilities and their application within specific work contexts, resulting in practical and useful ways of integrating AI across a variety of workplaces. Through this process, we enable effective AI integration while advancing stakeholder-centered AI literacy.

\newpage
\bibliographystyle{ACM-Reference-Format}
\bibliography{main}

\end{document}